\documentclass{aastex}
\usepackage{emulateapj5}

\usepackage{textcomp}
\usepackage{epsfig, epstopdf}
\usepackage{longtable,lscape}
\usepackage{amssymb,amsmath}

\newcommand{\simgt}{\lower 2pt \hbox{$\, \buildrel {\scriptstyle >}\over {\scriptstyle\sim}\,$}}
\newcommand{\simlt}{\lower 2pt \hbox{$\, \buildrel {\scriptstyle <}\over {\scriptstyle\sim}\,$}}

\newcommand{\rxj}{RX~J1131$-$1231}

\slugcomment{Received 2012 April 16}
\shorttitle{Energy Dependent X-ray Microlensing}
\shortauthors{CHARTAS ET AL.}

\begin{document}

\def\sarc{$^{\prime\prime}\!\!.$}
\def\arcsec{$^{\prime\prime}$}
\def\beginrefer{\section*{References}%
\begin{quotation}\mbox{}\par}
\def\refer#1\par{{\setlength{\parindent}{-\leftmargin}\indent#1\par}}
\def\endrefer{\end{quotation}}

\title{Revealing the Structure of an Accretion Disk Through Energy Dependent X-ray Microlensing}

\author{G. Chartas\altaffilmark{1}, C. S. Kochanek\altaffilmark{2}, X. Dai\altaffilmark{3}, D. Moore\altaffilmark{1},  A. M. Mosquera\altaffilmark{2}, and J.~A. Blackburne\altaffilmark{2}}

\altaffiltext{1}{Department of Physics and Astronomy, College of Charleston, Charleston, SC, 29424, USA, chartasg@cofc.edu}

\altaffiltext{2}{Department of Astronomy and the Center for Cosmology and Astroparticle Physics, The Ohio State
University, Columbus, OH 43210, USA}

\altaffiltext{2}{Homer L. Dodge Department of Physics and Astronomy, The University of Oklahoma, Norman, OK, 73019, USA}

\begin{abstract}

\noindent
We present results from monitoring observations of the gravitationally lensed quasar RX~J1131$-$1231
performed with the {\sl Chandra X-ray Observatory}.
The X-ray observations were planned with relatively long exposures that allowed a search for energy-dependent 
microlensing in the soft~(0.2-2~keV) and hard~(2-10~keV) light curves of the images of RX~J1131$-$1231. 
We detect significant microlensing in the X-ray light-curves of images A and D, and energy-dependent microlensing of image D. 
The magnification of the soft band appears to be larger than that in the hard band by a factor of  $\sim$ 1.3 
when image D becomes more magnified. This can be explained by the difference between a compact, 
softer-spectrum corona that is producing a more extended, harder
spectrum reflection component off the disk. This is supported by the evolution of the fluorescent iron line
in image D over three consecutive time-averaged phases of the light curve.
In the first period, a Fe line at E = 6.36$_{-0.16}^{+0.13}$~keV is detected (at $  > $ 99\% confidence). In the second period,  
two Fe lines are detected, one at E = 5.47$_{-0.08}^{+0.06}$~keV (detected at $  > $ 99\% confidence)
and another at E = 6.02$_{-0.07}^{+0.09}$~keV (marginally detected  at $  > $ 90\% confidence),
and in the third period, a broadened Fe line at 6.42$_{-0.15}^{+0.19}$~keV is detected (at $  > $ 99\% confidence). 
This evolution of the Fe line profile during the microlensing event 
is consistent with the line distortion expected when a caustic passes over
the inner disk where the shape of the fluorescent Fe line is distorted by General Relativistic and Doppler effects.
\end{abstract}

\keywords{galaxies: active --- quasars: 
individual~(RX~J~1131--1231) --- X-rays: galaxies --- gravitational lensing} 

\section{INTRODUCTION}
Some of  the most fascinating properties of quasars arise from the strong gravitational field
that dominates over all other forces near the supermassive black hole.
Under the effects of strong gravity and rapid rotation, the energies and kinematics of X-ray photons
emitted at a few gravitational radii are modified significantly.
Observations of the X-ray spectra of quasars thus have the potential of testing the theory of General Relativity in a 
region of strong gravity, constraining the structure of the emitting region and  
providing estimates of the spin of the black hole (e.g., Fabian et al 1989; Laor 1991).

A promising  technique for indirectly mapping the accretion disk around the black hole,
and thus constraining the properties of the black hole, relies on 
monitoring variations in the profiles of Fe fluorescence lines originating from the inner parts of the disk (e.g., Fabian et al 1989; Laor 1991). 
The iron fluorescence line is produced by X-rays from a hot corona irradiating 
the accretion disk. The line typically contains a narrow and a broad component. 
The narrow component is thought to
originate from material in the broad-line region and/or a molecular torus.
The broad component  most likely originates from accreting material closer to the black hole.
The iron line method has been most successful in studies of Galactic black-hole binaries 
where the X-rays are mostly thermal emission from the accretion disk (e.g., Done, Gierli{\'n}ski,  \& Kubota \ 2007 and references therein).
For the cooler, quasar disks, the X-rays are dominated by the non-thermal
emission from the corona rather than direct emission from the disk, making it more difficult to understand and model.

The broad line profiles are determined by a combination of the Doppler, special and general 
relativistic (GR) effects (e.g., Fabian et al 1989; Laor 1991), while their variations are caused by fluctuations in the 
X-ray continuum that drives the line emission (Young \& Reynolds 2000). Modeling of the distorted iron line 
can provide information about the disk geometry, inclination 
and the spin of the black hole (e.g.,  Fabian et al. 2000; Reynolds \& Nowak 2003).
The accuracy of the Fe line method for constraining the spin of a black hole from the broadened red wing of the Fe line profile
has been questioned recently (e.g., Miller et al 2008; Sim et al. 2010).
In particular, simulations by Sim et al. (2010) show that the Fe  line can develop 
a weak, red-skewed line wing as a result of Compton scattering in an outflow.
Separating the broadening of the Fe line  by GR effects near the inner edge of the disk from that produced by scattering  
in a wind may be difficult, especially with the limitations of low resolution X-ray spectra.

An independent method of probing the structure of quasars relies on gravitational microlensing
(e.g., Paczynski 1986; Wambsganss 1990).
Stars in the foreground lens galaxy produce complex microlensing magnification maps
that pass over the quasar. The characteristic size of a magnification caustic produced by a single star is
referred to as the Einstein radius. The strength of the magnification depends on the relative size
of the emitting region and the Einstein radius, with more compact sources having higher microlensing variability amplitudes.

Microlensing observations constrain the sizes of accretion disks at optical wavelengths
(Pooley et al. 2007; Anguita et al. 2008; Morgan et al. 2010; Jimenez-Vicente et a. 2012),  the sizes of quasar coronae 
at X-ray energies (Chartas et al. 2002, 2007, 2009; Pooley et al. 2007; Morgan et al., 2008; Dai et al. 2010; Blackburne et al. 2011b)
and the dependence of the accretion disk size with wavelength (Poindexter et al. 2008; Bate et al. 2008;  Eigenbrod et al. 2008;  Floyd et al. 2009;  
Morgan et al. 2010; Blackburne et al. 2011a; Mosquera et al. 2009, 2011b).
A study of microlensing variability in a sample of eleven microlensed quasars performed by Morgan et al. (2010)
showed that the radius of the accretion disk scales as $M^{2/3}_{BH}$ and the sizes estimated from  microlensing 
are larger than predicted by thin disk theory for the typical Eddington ratios, ($L_{\rm Bol}/L_{\rm Edd}$ $\sim$ 1/3), of quasars.

We recently presented the results of a six epoch monitoring campaign of the $z_{\rm s}$=0.658 quasar \rxj\ (Chartas et al. 2009; Dai et al. 2010).
In these studies we found  significant flux variability in all four images 
and that the saddle point image A was strongly microlensed (also see Blackburne et al. 2006; Pooley et al. 2007).
In Dai et al. (2010) we modeled the X-ray and rest-frame UV  light curves of \rxj\ in detail using the
technique introduced in Kochanek  (2004),  and found that the 
X-ray emission in \rxj\ comes from a region smaller than $\sim 10 r_g$
while  the observed optical emission region has a size of  $\sim 70 r_g$, where
 $r_{\rm g} = GM_{\rm BH}/c^{2}$ is the gravitational radius of the black hole and \rxj\ is estimated 
 to have $M_{\rm BH} = 1.3\times 10^{8} M_\odot$ based on the $H_{\beta}$ line width (Peng et a. 2006).

Here we present the results from the continuation of our X-ray and optical monitoring 
campaigns for \rxj.  The X-ray exposure times in the new epochs
have been increased to provide sufficient signal-to-noise spectra to search for X-ray 
energy dependent microlensing.
In \S 2 we present the analysis of the X-ray observations, in \S 3.1 we discuss 
the detection of microlensing events in the light curves of images A and D,
in \S 3.2 we present spectra possibly showing the evolution of the Fe line as a caustic crosses the black hole, and 
in \S 3.3 we present constraints on quasar structure assuming the detected Fe line energies are correct. 
Throughout this paper we adopt a flat $\Lambda$ cosmology with 
$H_{0}$ = 70~km~s$^{-1}$~Mpc$^{-1}$,  $\Omega_{\rm \Lambda}$ = 0.7, and  $\Omega_{\rm M}$ = 0.3.

\section{X-RAY OBSERVATIONS AND DATA ANALYSIS}

\rxj\ was observed with the Advanced CCD Imaging Spectrometer (ACIS; Garmire 2003) on board 
the {\sl Chandra X-ray Observatory} (hereafter {\sl Chandra})
29 times between 2004 April 12 and 2011 February 25.
Results from the analysis of the first six observations of \rxj\ were presented in 
Chartas et al. (2009) and Dai et al. (2010). Here we describe the data analysis for the remaining 23 observations, 
but we will use the results from all 29 observations in our microlensing analysis.

The {\sl Chandra} observations of \rxj\ were analyzed using the standard software CIAO 4.3 
provided by the {\sl Chandra X-ray Center} (CXC). 
A log of the observations that includes observation dates, observational identification numbers, exposure times, ACIS frame time 
and the observed 0.2--10~keV counts is presented in Table 1.
We used standard CXC threads to screen the data for status, grade, and time intervals of acceptable aspect solution and background levels. 
To improve the spatial resolution we employed the sub-pixel resolution technique developed by Tsunemi et al. (2001).

We estimate the X-ray counts of the images in the total (0.2--10~keV), soft (0.2--2~keV) and 
hard (2-10~keV) bands. The total counts were corrected for pile-up for 
images A and B, and pile-up is unimportant for images C and D.
The hard and soft bands for images A and B were not corrected for pile-up because of the difficulty in estimating energy-dependent pile-up corrections at the observed count rates. 
Pile-up is an instrumental effect that occurs when two or more X-ray photons strike individual or neighboring CCD pixels within 
one frame time and can cause spectral and point-spread-function (PSF) distortions, grade migration, 
and an apparent hardening of the energy spectra.
We used the  forward spectral-fitting tool LYNX
(Chartas et al. 2000) to estimate the fraction of events lost due to the pile-up
effect. These corrections affect image A (up to 53\% in epoch 23) and B (up to 22\% in epoch 18).
Pooley et al. (2012) in their analysis of the epochs from 2004 April 12 to 2008 April 11 did not include these corrections, which
may explain some of the epoch-dependent results in their analysis.

We fit the lensed images simultaneously using the relative astrometry derived from {\sl HST} images (Morgan et al. 2006)
and \verb+MARX+ models (Wise et al. 1997) of the PSF.
We fit the data using 0\sarc0246 bins (compared to the 0\sarc491 ACIS pixel scale)
by minimizing the $C$-statistic (Cash 1979) between the observed and model images.
The PSF fitting method provided the best fit value and errors for the number of counts in each image, $N_{\rm i,psf}$ and 
$\sigma_{\rm i,psf}$, respectively. We define as \(N_{\rm ABCD,psf}  = \sum_{i} N_{\rm i,psf} \)
to be the total counts of all images from the PSF fitting method.

While the PSF models should correctly recover the relative fluxes, errors in the PSF
models may bias the total fluxes. We corrected for this by renormalizing the 
total counts to match the total counts found in a 5\arcsec\ radius region
centered on the lens with a background correction based on an annulus from 
7\sarc5  to 50\arcsec\ around the lens. Specifically, the counts from each image as determined from 
the PSF fitting method were renormalized to $N'_{\rm i,psf}$ = R$\times$$N_{\rm i,psf}$, where 
R = $N_{\rm ABCD,ap}$/$N_{\rm ABCD,PSF}$ and $N_{\rm ABCD,ap}$ is the total counts from the aperture method. 
Table 2 lists the exposure times and the $N'_{\rm i,psf}$ counts of 
images C and D  in the soft  and hard bands.

Figure 1 shows the stacked image of all observations of \rxj\ listed in Table 1,
with a total effective exposure time of  $\sim$ 333.6~ks. 
To reduce background contamination and to enhance possible soft extended X-ray emission we
filtered the data to include only photons with energies from 0.4 and 3.0~keV.
The image was binned with a bin size of 0.5 arcsec and adaptively smoothed with the CSMOOTH tool developed by 
Ebeling et al. (2000). CSMOOTH smooths a two-dimensional image with a circular Gaussian kernal of varying radius.
We find no emission from a possible cluster of galaxies in the field with a 3~$\sigma$ upper limit on the 0.2--2~keV  luminosity  of
6.1 $\times$ 10$^{41}$ ergs~s$^{-1}$, assuming a cluster with a temperature of 
1.5 keV at the lens redshift, an abundance of 0.3 solar and an extraction radius of 30 arcsec.

The stacked image of \rxj\ does not show any additional lensed images.
To obtain quantitative limits on possible additional images located at the center of  \rxj\ 
we extracted the 0.2--10~keV counts within a 0.4~arcsec circle centered 
on the four images and the lens.
The backgrounds in these extraction regions are dominated by the contamination
from the images rather than the instrumental or cosmic backgrounds.
To estimate the fractional contamination per image we used the 0.2--10~keV counts in 
0.4~arcsec circular apertures placed 1 and 2~arcsec North of image B in \rxj.
Because the summed image averages over many spacecraft orientations, the averaged PSFs are largely
free of angular structure.
The detected 0.2--10~keV counts in the 0.4 arcsec aperture centered on the lens are
consistent with the estimated background and contamination from the bright images.
Combining all the data we obtain a 3$\sigma$ upper limit on the 0.2--10~keV flux
of any central, odd image of $4.7\times 10^{-15}$ ergs~s$^{-1}$.   
This limit corresponds to a 3$\sigma$ limit on the
flux ratio of a factor of $\sim$ 135 relative to the average flux of image C during period 2. 
While this is a strong limit on the flux of a central image, it is not yet strong enough to place 
interesting constraints on the central density of the lens (see Keeton 2005).

In Figure 2 we show the 0.2-10 keV light-curves of images A, B, C and D.
Images A and D show significant uncorrelated flux variability  suggesting that these images are
also significantly affected by microlensing.
The light-curves have been shifted to account for the estimated time delays 
between the images of \rxj\ (Chantry et al. 2010).
The lower amplitude variability seen in images B and C is strongly correlated, suggesting that these images are 
less affected by microlensing.

In Figure 3 we show the observed X-ray and optical flux ratios of D/B, A/B, and C/B.
The $R$-band  monitoring data shown in Figure 3 are obtained with the SMARTS 1.3m and ANDICAM (Depoy et a. 2003).
They are reduced using the standard procedures and modeled as described in Kochanek et al. (2006).
The time delays between images are considerably shorter than the time scales of the two major 
magnification events seen in images A and D and the intrinsic brightness of the source is fairly constant
over the observational period, as seen in the light-curves of images C and B. 
We therefore do not expect the observed flux ratios to be significantly affected by intrinsic variability
over the period of the strong magnification events in images A and D.

The total X-ray band flux ratio C/B is relatively constant at 0.29 $\pm$ 0.03, supporting our
hypothesis that the microlensing magnifications of images B and C are relatively constant
during our new observations. It also indicates that the intrinsic flux of the quasar does not
vary by more than 28\% during the observing period.
Specifically, the ratios of the standard deviations to the mean total band X-ray count rates 
of images A, B, C and D are 0.58, 0.28, 0.24, and 0.75, respectively.
The large flux variations detected in images A and D are uncorrelated with those of C and B and are 
therefore not intrinsic to the quasar but must be the result of microlensing.

Image A is significantly and image  B is mildly piled-up, and, as a result, their spectra  
are expected to appear flatter in energy and distorted. We therefore focus the spectral analysis on images 
C and D.  We fit the {\sl Chandra}  spectra of images C and D of \rxj\  
for each epoch with a model that consists  of a power law with neutral intrinsic absorption at $z_{\rm s}$ = 0.658. 
We also included Galactic absorption due to neutral gas with a column 
density of N$_{H}$ = 3.6 $\times$ 10$^{20}$ cm$^{-2}$ (Dickey \& Lockman 1990).
The results of these fits are presented in Table 3. As shown in Table 3, the values of the photon indices 
and X-ray fluxes of image C do not show any significant 
variations.  Significant spectral variability is detected in image D that is likely associated 
with energy-dependent microlensing. The energy dependent microlensing interpretation is discussed in more detail in the section \S 3. 
The spectral fits also indicate that there is no significant absorption at the lens or source redshifts.

\section{DISCUSSION}

\subsection{Microlensing events  \rxj}

Significant deviations of the flux ratios from a constant value over time are suggestive of microlensing 
since the  intrinsic variability of the quasar cancels in a flux ratio up to our ability to correct for time delays.
We observe significant variability in the ratios D/B, D/C, A/B and A/C but not in C/B suggesting that only 
images A and D show significant microlensing activity during our new observations (e.g., see Figure 3).
Specifically, we find that the total band flux ratio A/B deviates significantly from a constant  between days 4800
and 5600, a period of about 800 days. The total flux ratio D/B deviates from a constant for a period of about 1,100 days.

Microlensing variability occurs when the complex pattern of caustics produced by stars in the lens 
moves across the source plane. 
The characteristic scale of these patterns is the Einstein radius of

\begin{equation}
{R_{\rm E} =  \left[  {{4G \langle M \rangle }\over{c^{2}}}  {D_{\rm OS} {D_{\rm LS}}\over{D_{\rm OL} }}  \right ]^{1/2}} = 2.5 \times 10^{16}\left( {{ \langle M \rangle }\over{0.3M_{\odot}} }\right)^{1/2}~{\rm cm}
\end{equation}

\noindent
where $\langle M \rangle$ is the mean mass of the lensing stars, D is the angular diameter distance, and the subscripts L, S, and O refer to the lens, source, and observer, respectively. 
Emission regions with sizes significantly larger than $R_{\rm E}$ will 
be affected little by microlensing, whereas emission regions significantly smaller than $R_{\rm E}$
can be strongly magnified.
Due to the combined motions of the observer, lens, source, and stars in the lens galaxy, the source moves relative
to the magnification patterns at an effective velocity of order 700 km~s$^{-1}$ for \rxj\ (Mosquera \& Kochanek 2011a).
The time scale to sample a significantly different microlensing region is the Einstein crossing time of 
$t_{\rm E}$ = $R_{\rm E}$/$v_{\rm c}$  $\approx$ 11~years, which is roughly the temporal
baseline of the X-ray data. Microlensing variability can be seen on much shorter time scales corresponding to the time scale needed to
move by the size of the source, $t_{\rm S}$ = $R_{\rm S}$/$v_{\rm c}$.
For the optical ($R_{\rm O}$ = 1.3 $\times$ 10$^{15}$cm)
and X-ray ($R_{\rm X}$ = 2.3 $\times$ 10$^{14}$cm) sizes we found in Dai et al.  (2010),
the source crossing times are of the order $t_{\rm O}$ =  $R_{\rm O}$/$v_{\rm c}$ = 0.6~years
and $t_{\rm X}$ =  $R_{\rm X}$/$v_{\rm c}$ = 0.1~years.

Figure 3 shows these time scales in comparison to the changes in the flux ratios, and it is striking that the 
apparent time scales in the X-ray and optical A/B and D/B flux ratios closely match the predictions from the earlier size estimates.
For comparison, the estimated black hole mass of 6 $\times$ 10$^{7}$ M$_{\odot}$ (Peng et al. 2006)
corresponds to a Schwarzschild radius of  $R_{S}$ = 1.8 $\times$ 10$^{13}$~cm.
Microlensing magnification patterns have, however, strong spatial correlations with dense networks of
high magnification caustic regions separated by broad valleys with little activity.
Thus, an image showing strong microlensing variability is likely to continue to do so for a time scale
of order $t_{\rm E}$, while one without such variability is also likely to continue in such a state. 
Previous studies of \rxj\ suggest that images A and D are ``active"  and images C and B are not.

\subsection{Energy Dependent Microlensing in \rxj}

Figure 4 shows the soft (0.2-2~keV) and hard (2-10~keV) flux ratios D/B and D/C.
The total, soft, and hard band flux ratios D/C and D/B are almost constant with time between days  $\sim$ 4000 $-$ 4500.
After this period the flux ratios increase, indicating the onset of significant microlensing magnification of image D.
The soft and hard band flux ratios begin to differ at a modified Julian date of $\sim$ 4,700 days, just after an increase in the 
microlensing magnification of image D.
Specifically, the soft band flux ratios  D/B  and D/C are systematically larger than the hard band ratios by a constant  
factor of 1.3 $\pm$ 0.1 for a period of about 600~days.
We therefore associate these differences of the soft and hard band flux ratios D/B and D/C with an energy dependent microlensing event in image D.
Image A is significantly affected by pile-up and any estimate of the soft and hard band fluxes
of this image is very uncertain. We have therefore not attempted to correct the soft and hard band
fluxes of image A, but only its total band flux.

It is possible to explain the energy dependent behavior of image D as
a caustic approaching and traversing the accretion disk. The X-ray emission is sufficiently compact that its 
microlensing variability is probably dominated by the effects of individual caustics 
(although there will be exceptions).
We assume the X-ray spectrum is a combination of a direct X-ray component from a hot corona that is centered on the black hole 
and a more extended reflected component from the accretion disk.
As shown in Figure 5, adapted from Reynolds et al. (1998),  the reflected X-ray component is harder than the 
coronal component.
The ratio of reflected-to-direct emission is larger in the hard band than in the the soft band,

\begin{equation}
\left(\frac{f_{Refl}}{f_{Direct}}\right)_{Hard} >  \left(\frac{f_{Refl}}{f_{Direct}}\right)_{Soft},
\end{equation}

\noindent
where $f_{\rm Refl}$ is the flux of reflected emission off the accretion disk and $f_{\rm Direct}$ is the flux of direct emission from the hot corona.

When the central source is far away from a caustic, the two components have similar magnifications and 
the observed spectrum is dominated by direct emission
from the corona with a smaller contribution from
reflected disk emission. As the caustic approaches the center of the black hole, 
the more extended reflected X-ray component is magnified first.
The X-ray  spectrum at this point is still dominated by the direct component 
and the ratio of soft-to-hard band is not expected to change significantly.
As the source approaches the caustic, the more compact corona becomes strongly magnified.
Since the coronal spectrum  is softer than the reflected, the
soft band will be more magnified than the hard band 
as the source crosses and passes the caustic.

We can use the evolution of the X-ray spectrum of image D to test 
our explanation.
To improve the S/N we stacked the X-ray spectra of images C and D in three periods.
Period 1 includes epochs 7 though 16, period 2 includes epochs 17 through 22,
and period 3 includes epochs  23 through 29.
In Figure 6 we overlay the stacked spectra of images C and D for the three periods
to show the spectral evolution of the continuum and the lines. The stacked spectra of image D are also displayed
in separate panels in Figure 7 to  clearly show the iron lines.
In Figure 8 we show the 68\%, 90\% and 99\% confidence contours  of the flux normalizations of the
Fe lines detected in images C and D versus their  energies during the three periods.

During period 1the D/C flux ratios remain constant given the uncertainties, indicating that 
the microlensing magnifications are roughly constant.
A significant Fe K line is detected at an energy of 6.36$_{-0.16}^{+0.13}$~keV
that is unresolved in width. The 0.2-10~keV flux in the entire spectrum of image D for period 1  
is 2.1 $\times$ 10$^{-13}$~erg~s$^{-1}$~cm$^{-2}$,
the flux in the Fe line is  4.2 $\times$ 10$^{-15}$~erg~s$^{-1}$~cm$^{-2}$ and its rest-frame equivalent width is $W_{\rm E} = 398_{-398}^{+414}$~eV.
During period 2, the D/C flux ratios  begin to increase, indicating a change in the magnification
of image D, but no energy dependence is detected in the flux ratios.
A significant Fe K line is detected at an energy of 5.47$_{-0.08}^{+0.06}$~keV (unresolved in width)
and a second line is marginally detected at an energy of 6.02$_{-0.07}^{+0.09}$~keV (unresolved in width).
The 0.2-10~keV flux in the entire spectrum of image D for period 2 is 2.7 $\times$ 10$^{-13}$~erg~s$^{-1}$~cm$^{-2}$
and the fluxes (rest-frame equivalent widths) in the 5.48~keV and  6.02~keV lines are  3.4 $\times$ 10$^{-15}$~erg~s$^{-1}$~cm$^{-2}$
($W_{\rm E} = 210_{-175}^{+160}$~eV)
and  3.0 $\times$ 10$^{-15}$~erg~s$^{-1}$~cm$^{-2}$($W_{\rm E} = 200_{-185}^{+180}$~eV), respectively.
During period 3, the D/C flux ratios  are significantly enhanced and 
the soft band flux ratios are larger than the hard band ratios by a factor of 1.3 $\pm$ 0.1.
A significant broad Fe K line is detected at an energy of 6.42$_{-0.15}^{+0.19}$~keV
with a width of 220$_{-170}^{+250}$~eV. The 0.2-10~keV flux in the entire spectrum of image D for period 3 is 7.5 $\times$ 10$^{-13}$~erg~s$^{-1}$~cm$^{-2}$
and the flux (rest-frame equivalent width) in the Fe line is 1.2$\times$ 10$^{-14}$~erg~s$^{-1}$~cm$^{-2}$($W_{\rm E} = 290_{-120}^{+130}$~eV).
The spectrum of image C is essentially unchanged over the three periods, as shown in Figure 6. In Figure 9 we show the stacked 
spectrum of image C from combining all epochs.
A significant Fe K line is detected at an energy of 6.37$_{-0.08}^{+0.07}$~keV.
The average 0.2-10~keV flux in the combined spectrum of image C over the three periods is 6.1 $\times$ 10$^{-13}$~erg~s$^{-1}$~cm$^{-2}$
and the flux (rest-frame equivalent width) in the Fe line is 4.5$\times$ 10$^{-15}$~erg~s$^{-1}$~cm$^{-2}$($W_{\rm E} = 140_{-70}^{+80}$~eV).

The X-ray data imply that the ratio of the total flux (corona dominated) to Fe line flux increases from Period 2 to 3, 
which would imply that the spectrum of D would become softer from Period 2 to 3
since the coronal spectrum is softer than the reflected one. In Figure 10 we show the ratios of the
spectra and the best-fit absorbed power-law plus line models of images D and C.
Specifically, we plot the observed and modeled ratios $D(E)_{\rm i}/C(E)$, 
where, $D_{\rm i}(E)$ is the flux density for image D in period $i$ and $C(E)$ is the flux density  for
image C in all three periods. The three periods are shown in Figure 4.
We detect a significant softening of the spectrum during the onset of the 
strong microlensing event  in image D that occurs between Periods 2 and 3. 
We note that the softening of the spectrum of an image during a microlensing event is opposite to what
we observed in Q2237+0305 (Chen et al. 2011).

\subsection{General and Special Relativistic Effects}

Assuming both detections of Fe line components in image D during period 2 are correct, we can
interpret them as the result of distortions of the fluorescent Fe line due to 
general relativistic (GR) and special relativistic (SR) Doppler effects.
Detailed models are likely unwarranted given the present data, so we will focus on a semi-quantitative outline
of what we have observed. Assuming that the Fe~K$\alpha$ emission arises from the disk,
the energy shift of the midpoint of the two lines can be used to constrain the 
radius of the emitting material or equivalently the distance of the caustic from the 
black hole. The splitting of the lines is likely the result of the Doppler effect arising
from an approaching and receding part of the disk.
Both lines are expected to be redshifted by the GR effect but also 
for large velocities special relativity predicts a redshift of both lines.

The gravitational redshift  of the Fe line is

\begin{equation}
{\Delta}E_{\rm GR} = {E_{0}}\left[\left( 1- 2\left(\frac{r_{\rm g}}{r_{\rm em}}  \right)  \right)^{0.5}-1      \right],
\end{equation}

\noindent
where $r_{\rm g}$ is the gravitational radius, $r_{\rm em}$ is the radius of the line emitting region and $E_{0}$ = 6.4~keV.
The redshift due to special relativistic effects (relativistic Doppler) 
of the midpoint of the two lines is

\begin{equation}
{\Delta}E_{\rm SR} = {E_{0}}\left[\left( 1- \left(\frac{r_{\rm g}}{r_{\rm em}}  \right)  \right)^{0.5}-1      \right].
\end{equation}

\noindent
Here ${\Delta}E_{SR}$ only includes the redshift
of the midpoint of the two lines and not their splitting. We have assumed Keplerian velocities
with $(v_{k}/c)^{2} = r_{g}/r_{em}$ for simplicity. The predicted total shift of the midpoint between the two lines from $E_{0}$ = 6.4~keV is 
${\Delta}E_{Total} =  {\Delta}E_{GR} + {\Delta}E_{SR}$. 
We observe a total energy shift ${\Delta}E_{Total} = 0.65$~keV.
This energy shift implies (based on eqs. [3] and [4]) that $r_{em} \sim 15r_{g}$ and the corresponding Keplerian velocity
at this radius is 0.26$c$. 
The estimated value of the Keplerian velocity, $v_{\rm k}$, combined with the observed separation of the two lines provides a
constraint on the inclination angle, $i_{incl}$,  of the disk. Specifically, the energy separation of the two lines is

\begin{equation}
{\Delta}E_{\rm DOP} = {2E_{0} \left(\frac{v_{k}}{c}  \right)\cos\theta}\left[{\gamma}\left( 1- \left(\frac{v_{k}}{c}  \right)^{2}\cos^{2}\theta  \right)      \right]^{-1}  
\end{equation}

\noindent
where, $\gamma$ is the Lorentz factor and $i_{incl} = 90^{\circ}  - \theta$.
The observed line separation of 0.55 $\pm$ 0.10~keV implies (from  eq. [5]) an inclination angle
$i_{incl} = 10_{-2}^{+2}$ degrees.

The direction of motion of the caustic with respect to the projected disk axis
will affect the relative amplitude of the lines.
This may explain why the Fe line from the receding material at E = 5.5~keV
is brighter than the line at E = 6~keV from the approaching material.
The redshifted lines are narrow and not broad as typically
observed in Seyfert galaxies. This is likely the result of
the caustic selectively magnifying
emission within a narrow radius as it approaches the black hole.

As we showed in Figure 6 the continuum and the line emission are relatively unchanged between epochs in the non-microlensed image C.These spectra confirm that the quasar is not significantly 
showing any intrinsic  spectral variability between observing periods.
The detected shift of the Fe line in image D of \rxj\ is therefore not the result
of an intrinsic phenomenon such as a rotating hotspot.
This reinforces our conclusion that the shift in the energy of the iron line
in the microlensed image D during period 2 is the result of selective magnification of material closer in to the black hole
as the caustic crosses the accretion disk.

We have observed time-dependent X-ray microlensing in all seven systems
examined in our {\sl Chandra} monitoring program (Chen et al. 2012).  We see energy dependent
effects in five systems, although in all cases the differences are modest  in the sense that the 
differences between the hard and soft bands are far smaller than the differences between 
the X-ray and optical microlensing.  
Thus, while the hard and soft X-ray emission regions may have different
average sizes, those differences are not large.  Similarly, all our
analyses to date indicate that the X-ray emission region has a size of
order $10r_{\rm g}$.  
Our upper limits on the X-ray sizes are robust, however, our lower limits are currently not well constrained.
An improvement on the lower limits 
requires stronger limits on the maximum
possible microlensing variability amplitude, information that is easily obtained with 
additional monitoring of these objects.
The observed changes in the Fe K$\alpha$
line structure implies that the Fe emission region of \rxj\ is 
compact, regardless of the details of our simple quantitative model.
For the Fe line study, the challenge is simply to obtain enough X-ray photons 
to fit the Fe line with more realistic models 
that can provide constraints on the structure of the disk and the properties of the black hole.
A strategy than can address both the lower bound and the Fe line  problems is
to observe lenses similar to \rxj\ that have a high probability of showing Fe line microlensing using
a denser time sampling of epochs that are individually long enough to
measure the hard and soft band fluxes accurately. The denser time sampling would solve the
minimum source size problem.  Co-adding the epochs would then provide a
study of the Fe K$\alpha$ line at the cadences we presently achieve
for the continuum flux ratios.

\acknowledgments
We acknowledge financial support from NASA via the Smithsonian Institution grants SAO GO0-11121A/B/C/D, SAO GO1-12139A/B/C,
and GO2-13132A/B/C.
CSK, JAB, and AMM  are supported by NSF grant AST-1009756.
GC thanks the inspirational environments offered by Baked and Kudo in Charleston SC, and 
also thanks Chris Fragile for insightful discussions.

\clearpage
\scriptsize
\begin{center}
\begin{tabular}{clccrlrrrr}
\multicolumn{10}{c}{TABLE 1}\\
\multicolumn{10}{c}{Log of Observations of Quasar \rxj\ } \\
 & & && & & && & \\ \hline\hline
         &        & & {\it Chandra}                                      & Exposure & & & &      \\
Epoch & Observation & JD${}^{a}$  &  Observation  &   Time   & $t_{\rm f}$${}^{b}$& $N_{\rm A}$$\tablenotemark{c}$ & $N_{\rm B}$$\tablenotemark{c}$ & $N_{\rm C}$$\tablenotemark{c}$ & $N_{\rm D}$ $\tablenotemark{c}$   \\
& Date           &     (days)        &          ID                &   (ks)       &(s)& counts   & counts & counts & counts   \\
&   &    &    &   &  &  & & &\\
\hline
1&2004 April 12             &3108           &  4814     & 10.0 &3.14&       425$_{-22}^{+22}$ & 2950$_{-54}^{+54}$ &  839$_{-29}^{+29}$ & 211$_{-15}^{+15}$  \\
2&2006 March 10          & 3805        &  6913    & 4.9    &0.741&     393$_{-20}^{+20}$  & 624$_{-25}^{+25}$  & 204$_{-14}^{+14}$ &  103$_{-10}^{+10}$    \\
3&2006 March 15           & 3810        &  6912    &  4.4   &0.741&       381$_{-20}^{+20}$  & 616$_{-25}^{+25}$ & 233$_{-15}^{+15}$  & 93$_{-10}^{+10}$  \\
4&2006 April 12               & 3838        &  6914    &  4.9   &0.741&     413$_{-20}^{+20}$   & 507$_{-23}^{+23}$  & 146$_{-12}^{+12}$   & 131$_{-12}^{+12}$  \\
5&2006 November 10    & 4050       &  6915     &  4.8  &0.741&     3708$_{-61}^{+61}$   & 1411$_{-38}^{+38}$ & 367$_{-19}^{+19}$   & 155$_{-13}^{+13}$  \\
6&2006 November 13    &4053        &  6916    &  4.8   &0.741&     3833$_{-62}^{+62}$  & 1618$_{-40}^{+40}$  & 415$_{-20}^{+20}$   & 115$_{-11}^{+11}$  \\
7&2006 December 17     & 4087&7786       & 4.88          &0.841& 3541$_{-99}^{+102}$   & 1443$_{-58}^{+59}$     &417$_{-28}^{+29}$      & 117$_{-13}^{+15}$\\
8&2007 January 01          &4102 &7785       & 4.70              &0.441& 2305$_{-74}^{+75}$    & 1082$_{-49}^{+51}$     & 312$_{-25}^{+26}$     & 108$_{-13}^{+15}$\\
9&2007 February 13        & 4145 &7787       & 4.71             &0.441& 2451$_{-78}^{+79}$    & 1116$_{-49}^{+51}$     & 301$_{-24}^{+25}$     & 169$_{-17}^{+18}$\\
10&2007 February 18      & 4150&7788       & 4.43           &0.441& 2232$_{-73}^{+75}$    &  950$_{-45}^{+47}$     & 252$_{-22}^{+23}$     & 115$_{-14}^{+15}$\\
11&2007 April 16             &4207  &7789       & 4.71                &0.441& 2328$_{-74}^{+74}$    & 1202$_{-53}^{+54}$     & 344$_{-26}^{+27}$     &  99$_{-12}^{+14}$\\
12&2007 April 25              &4216 &7790       & 4.70                &0.441& 2043$_{-70}^{+67}$    & 1063$_{-48}^{+50}$     & 363$_{-26}^{+28}$     & 142$_{-16}^{+17}$\\
13&2007 June 04             &4256 &7791       & 4.66              &0.441& 2079$_{-69}^{+70}$    & 1480$_{-59}^{+60}$     & 373$_{-27}^{+28}$     & 108$_{-13}^{+15}$\\
14&2007 June 11              &4263 &7792       & 4.68              &0.441& 2254$_{-72}^{+73}$    & 1466$_{-56}^{+58}$     & 337$_{-25}^{+26}$     & 129$_{-15}^{+16}$\\
15&2007 July 24               &4306 &7793       & 4.67                &0.441& 1958$_{-68}^{+69}$    & 1324$_{-55}^{+57}$     & 353$_{-25}^{+27}$     &  81$_{-11}^{+12}$\\
16&2007 July 30               & 4312&7794       & 4.67                &0.441& 2725$_{-79}^{+80}$    & 1844$_{-67}^{+69}$     & 496$_{-31}^{+32}$     &  100$_{-12}^{+13}$\\
17&2008 March 16           &4542 &9180       &14.32            &0.741& 5557$_{-117}^{+118}$  & 4347$_{-104}^{+105}$   &1337$_{-51}^{+53}$     & 351$_{-24}^{+25}$\\
18&2008 April 13             &4570 &9181       &14.35               &0.741& 8199$_{-147}^{+147}$  & 5654$_{-117}^{+118}$   &1453$_{-52}^{+52}$     & 377$_{-24}^{+25}$\\
19&2008 April 23             & 4580&9237       &14.31               &0.741& 6786$_{-130}^{+130}$  & 4927$_{-111}^{+112}$   &1279$_{-51}^{+51}$     & 232$_{-20}^{+21}$\\
20&2008 June 01            &4619 &9238       &14.24              &0.741& 4647$_{-106}^{+108}$  & 3252$_{-88}^{+90}$     & 878$_{-41}^{+43}$     & 463$_{-27}^{+28}$\\
21&2008 July 05              &4653    &9239       &14.28               &0.741& 5587$_{-118}^{+119}$  & 3584$_{-94}^{+95}$     &1001$_{-44}^{+44}$     & 635$_{-32}^{+33}$\\
22&2008 November 11  &4782&9240       &14.30       &0.741& 5135$_{-113}^{+115}$  & 3085$_{-85}^{+87}$     & 885$_{-42}^{+44}$     & 488$_{-29}^{+30}$\\
23&2009 November 28  &5164&11540       &27.52      &0.741&36024$_{-340}^{+342}$  & 7357$_{-128}^{+126}$   &2420$_{-68}^{+68}$     &3827$_{-81}^{+82}$\\
24&2010 February 09      &5237&11541       &25.62        &0.741& 26850$_{-290}^{+281}$  & 5814$_{-117}^{+117}$   &2059$_{-81}^{+86}$     &2437$_{-88}^{+79}$\\
25&2010 April 17              &5304&11542       &25.67             &0.741& 20935$_{-245}^{+246}$  & 6124$_{-119}^{+120}$   &1962$_{-62}^{+62}$     &2813$_{-69}^{+70}$\\
26&2010 June 25             &5373&11543       &24.62           &0.741&18521$_{-228}^{+230}$  & 5445$_{-111}^{+111}$   &1522$_{-54}^{+54}$     &1487$_{-51}^{+51}$\\
27&2010 November 11   &5512&11544       &25.56     &0.741& 27077$_{-467}^{+298}$  & 6316$_{-158}^{+120}$   &1821$_{-64}^{+57}$     &1730$_{-59}^{+56}$\\
28&2011 January 21        &5583&11545       &24.62        &0.741& 5689$_{-164}^{+153}$  & 3175$_{-123}^{+111}$   &1008$_{-53}^{+47}$     & 982$_{-46}^{+47}$\\
29&2011 February 2         &5618& 12833  &  13.61            &0.741& $5412_{-165}^{+159}$ &   $4375_{-128}^{+131}$ &   1161$_{-52}^{+52}$ &     $821_{-41}^{+44}$ \\

\hline \hline
\end{tabular}
\end{center}
${}^{a}$ Julian Date $-$ 245000.
${}^{b}$ {ACIS frame-time}. 
${}^{c}${Background-subtracted source counts for events with energies in the 0.2--10~keV band.
The counts for images A and B  of \rxj\ are corrected for pile-up.}\\

\clearpage
\scriptsize
\begin{center}
\begin{tabular}{lrrrr}
\multicolumn{5}{c}{TABLE 2}\\
\multicolumn{5}{c}{Soft (0.2-2~keV) and Hard (2-10~keV) counts of images C and D}\\
 & & & & \\ \hline\hline
          {}                              & Soft & Soft & Hard &  Hard    \\
 Observation       & $N_{\rm C}$ & $N_{\rm D}$ & $N_{\rm C}$ & $N_{\rm D}$ \\
 Date                         & counts   & counts & counts & counts   \\
    &       &  &  &  \\
\hline
2004 April 12   	           &            662$_{-26}^{+26}$& 172$_{-13}^{+13}$ &177$_{-13}^{+13}$  & 39$_{-6}^{+6}$ \\
2006 March 10           &               153$_{-12}^{+12}$& 77$_{-9}^{+9}$ &50$_{-7}^{+7}$ &  24$_{-5}^{+5}$ \\
2006 March 15	          &               172$_{-13}^{+13}$ &72$_{-8}^{+9}$ & 61$_{-8}^{+8}$  & 22$_{-5}^{+5}$ \\
2006 April 12 	          &               106$_{-10}^{+10}$& 103$_{-10}^{+10}$&40$_{-6}^{+6}$ &31$_{-6}^{+6}$ \\
2006 November 10   &               286$_{-17}^{+17}$& 117$_{-13}^{+13}$& 81$_{-9}^{+9}$ &38$_{-6}^{+6}$ \\
2006 November 13 	 &              320$_{-18}^{+18}$&98$_{-10}^{+10}$ & 96$_{-10}^{+10}$& 17$_{-4}^{+4}$ \\
2006 December 17    &              338$_{-19}^{+19}$ &     100$_{-9}^{+10}$ &       90$_{-10}^{+10}$ &     20$_{-4}^{+5}$ \\
2007 January 01        &              260$_{-17}^{+18}$ &      89$_{-9}^{+9}$ &        55$_{-8}^{+8}$ &       20$_{-4}^{+5}$ \\
2007 February 13       &             249$_{-16}^{+17}$ &     144$_{-12}^{+12}$ &      58$_{-8}^{+9}$ &       29$_{-5}^{+6}$ \\
2007 February 18        &              197$_{-14}^{+15}$ &      94$_{-9}^{+9}$ &        58$_{-8}^{+8}$ &       22$_{-4}^{+5}$ \\
2007 April 16                &             270$_{-17}^{+17}$ &      84$_{-9}^{+9}$ &        79$_{-9}^{+10}$ &      16$_{-4}^{+4}$ \\
2007 April 25               &              272$_{-17}^{+18}$ &     116$_{-11}^{+11}$ &      95$_{-10}^{+11}$ &     27$_{-5}^{+6}$ \\
2007 June 04              &              300$_{-18}^{+19}$ &      83$_{-8}^{+9}$ &        75$_{-9}^{+10}$ &      26$_{-5}^{+6}$ \\
2007 June 11              &             282$_{-17}^{+18}$ &     101$_{-10}^{+10}$ &      59$_{-8}^{+9}$ &       29$_{-5}^{+6}$ \\
2007 July 24               &              285$_{-17}^{+18}$ &      64$_{-7}^{+8}$ &        72$_{-9}^{+9}$ &       18$_{-4}^{+5}$ \\
2007 July 30               &              392$_{-21}^{+21}$ &      84$_{-9}^{+9}$ &        106$_{-11}^{+12}$ &    17$_{-4}^{+4}$ \\
2008 March 16           &            1058$_{-34}^{+35}$ &     256$_{-15}^{+16}$ &      285$_{-18}^{+18}$ &    97$_{-9}^{+10}$ \\
2008 April 13              &            1187$_{-36}^{+36}$ &     308$_{-16}^{+17}$ &      292$_{-17}^{+18}$ &    76$_{-8}^{+9}$ \\
2008 April 23              &             972$_{-32}^{+32}$ &     169$_{-12}^{+13}$ &      317$_{-20}^{+20}$ &    65$_{-8}^{+8}$ \\
2008 June 01              &           672$_{-26}^{+27}$ &     364$_{-18}^{+19}$ &      217$_{-15}^{+16}$ &    105$_{-10}^{+10}$ \\
2008 July 05                &             811$_{-30}^{+29}$ &     492$_{-21}^{+21}$ &      202$_{-15}^{+15}$ &    151$_{-12}^{+12}$ \\
2008 November 11      &             652$_{-26}^{+26}$ &     388$_{-20}^{+20}$ &      245$_{-17}^{+19}$ &    107$_{-10}^{+10}$ \\
2009 November 28       &           1952$_{-46}^{+47}$ &    3221$_{-57}^{+57}$ &      629$_{-26}^{+26}$ &    866$_{-28}^{+29}$ \\
 2010 February 09         &           1588$_{-59}^{+64}$ &    2017$_{-69}^{+59}$ &      558$_{-25}^{+25}$ &    523$_{-23}^{+23}$ \\
 2010 April 17                &           1541$_{-42}^{+41}$ &    2224$_{-45}^{+46}$ &      484$_{-22}^{+23}$ &    681$_{-26}^{+27}$ \\
2010 June 25                &           1213$_{-36}^{+36}$ &    1158$_{-33}^{+33}$ &      368$_{-20}^{+20}$ &    388$_{-20}^{+20}$ \\
2010 November 11       &           1460$_{-45}^{+36}$ &    1359$_{-40}^{+36}$ &      453$_{-22}^{+23}$ &    459$_{-22}^{+23}$ \\
2011 January 21           &            767$_{-36}^{+30}$ &     725$_{-28}^{+28}$ &      251$_{-17}^{+18}$ &    268$_{-19}^{+20}$ \\
2011 February 2           &            922$_{-36}^{+36}$ &     619$_{-25}^{+27}$ &      250$_{-16}^{+17}$ &    209$_{-16}^{+18}$ \\
       
 \hline \hline
\end{tabular}
\end{center}

\clearpage
\scriptsize
\begin{center}
\begin{tabular}{lcccccc}
\multicolumn{7}{c}{TABLE 3}\\
\multicolumn{7}{c}{Spectral Analysis of Image C of \rxj}\\
  & & & & &&\\ \hline\hline
 Observation                        &      $\Gamma$    &  $N_{\rm H}$($z_{\rm s}$=0.658)$^{a}$  & $\chi$$^{2}$/$\nu$ &{\it P}($\chi$$^{2}$/$\nu$)$^{b}$&Soft Flux  & Hard Flux  \\
 Date   &    & (10$^{22}$~cm$^{-2}$)  &    & & \multicolumn{2}{c}{ (10$^{-13}$ ergs cm$^{-2}$ s$^{-1}$)} \\
\hline
2004 April 12  	&	1.70$_{-0.11}^{+0.19}$ &     0.0$_{-0.0}^{+0.06}$&    54/56      &  0.57    &   3.14   &3.79 \\
2006 March 10&	1.72$_{-0.26}^{+0.29}$&      0.0$_{-0.0}^{+0.17}$&    14/15   &  0.52    &   1.26   &2.31 \\
2006 March 15	&	1.50$_{-0.22}^{+0.35}$&      0.0$_{-0.0}^{+0.30}$&    18.1/16   &  0.36    &    1.84&2.95 \\
2006 April 12	&	1.51$_{-0.35}^{+0.51}$&      0.0$_{-0.0}^{+0.35}$&    7.4/9      &  0.60    &    1.22&1.53 \\
2006 November 10 &	1.93$_{-0.19}^{+0.21}$&      0.0$_{-0.0}^{+0.09}$&    34/27       &  0.16    &    3.36&2.84 \\
2006 November 13	&	1.99$_{-0.21}^{+0.24}$&       0.1$_{-0.1}^{+0.2}$&    27/32       &  0.70    &    3.33&3.68 \\
2006 December 17       &        1.89$_{-0.17}^{+0.18}$ &     0.0$_{-0.0}^{+0.09}$ &    13.5/20     &   0.86   &    2.74    &       3.11    \\
2007 January 01       &         1.90$_{-0.23}^{+0.30}$ &      0.0$_{-0.0}^{+0.18}$ &   21.6/14     &   0.09   &    1.70      &     2.27        \\
2007 February 13        &         1.85$_{-0.19}^{+0.24}$ &     0.0$_{-0.0}^{+0.12}$ &     17.5/14     &  0.23    &   1.90        &   2.35            \\
2007 February 18       &        1.79$_{-0.25}^{+0.29}$ &      0.0$_{-0.0}^{+0.13}$ &      16.5/12     &   0.17    &   1.76      &    2.30         \\
2007 April 16      &         1.77$_{-0.21}^{+0.31}$ &      0.0$_{-0.0}^{+0.15}$ &    10.9/15     &   0.76    &   2.02      &     2.90           \\
2007 April 25       &         1.73$_{-0.27}^{+0.32}$ &     0.0$_{-0.0}^{+0.20}$ &    13.7/16      &  0.63    &   1.92        &   3.54          \\
2007 June 04        &         1.89$_{-0.19}^{+0.27}$ &      0.0$_{-0.0}^{+0.15}$ &     10.7/16      & 0.83    &   2.31        &   2.69          \\
2007 June 11      &         2.13$_{-0.27}^{+0.34}$ &     0.0$_{-0.0}^{+0.26}$ &    16.7/15      &  0.34    &   2.08       &    1.97           \\
2007 July 24       &         1.82$_{-0.18}^{+0.28}$ &      0.0$_{-0.0}^{+0.18}$ &    14.0/18      &  0.73     &  2.32       &    3.22         \\
2007 July 30        &         1.89$_{-0.16}^{+0.17}$ &      0.0$_{-0.0}^{+0.07}$ &    16.8/24      &  0.86     &  3.28         &  3.80             \\
2008 March 16        &         1.79$_{-0.08}^{+0.09}$ &     0.0$_{-0.0}^{+0.04}$ &    73.4/70       & 0.37     &  2.89        &   4.05           \\
2008 April 13         &        1.98$_{-0.09}^{+0.09}$ &     0.0$_{-0.0}^{+0.03}$ &    58.5/73      &  0.89     &  3.26         &  3.28          \\
2008 April 23       &         1.88$_{-0.09}^{+0.06}$ &     0.0$_{-0.0}^{+0.06}$ &     65.3/63      &  0.40    &   2.69        &   3.24          \\
2008 June 01       &        1.79$_{-0.10}^{+0.11}$ &     0.0$_{-0.0}^{+0.04}$ &    55.5/45       & 0.14     &  1.78        &   2.48          \\
2008 July 05       &       1.93$_{-0.11}^{+0.11}$ &     0.0$_{-0.0}^{+0.03}$ &   62.3/52     & 0.16   &    2.32      &   2.51           \\
2008 November 11        &       1.69$_{-0.10}^{+0.12}$ &     0.0$_{-0.0}^{+0.06}$ &   42.4/46      & 0.62    &   1.75        &   2.90           \\
2009 November 28       &      1.68$_{-0.08}^{+0.09}$ &    0.03$_{-.03}^{+0.06}$ &   88.6/113   &   0.96    &   2.36       &    4.28           \\
 2010 February 09       &       1.73$_{-0.07}^{+0.10}$ &    0.0$_{-0.0}^{+0.07}$ &     71.6/88    &    0.89   &   2.03      &    3.20        \\
2010 April 17        &        1.65$_{-0.06}^{+0.08}$ &    0.0$_{-0.0}^{+0.05}$ &   109.2/106    & 0.40   &    2.34        &   4.14              \\
2010 June 25        &        1.71$_{-0.08}^{+0.08}$ &    0.0$_{-0.0}^{+0.03}$ &    71.2/79     &  0.72    &   1.88        &   2.96          \\
2010 November 11      &         1.69$_{-0.07}^{+0.07}$ &    0.0$_{-0.0}^{+0.04}$ &  91.2/94    &   0.56     &  2.16          & 3.57       \\
2011 January 21       &          1.63$_{-0.09}^{+0.11}$ &     0.0$_{-0.0}^{+0.06}$ &   54.6/54     &  0.45    &  1.18        &   2.17           \\
2011 February 2         &          1.78$_{-0.10}^{+0.15}$ &     0.0$_{-0.0}^{+0.07}$ &    61.4/54      & 0.23     &  2.29       &    3.35             \\
\hline \hline
\end{tabular}
\end{center}
${}^{a}$All model fits include fixed, Galactic absorption of $N_{\rm Gal,H}$  = 0.036 $\times$ 10$^{22}$~cm$^{-2}$ (Dickey \& Lockman 1990)
and intrinsic ($z$ = 0.658) absorption $N_{\rm H}$ set as a free parameter.
All errors are at 90\% confidence on one parameter. \\
${}^{b}$$P(\chi^2/{\nu})$ is the probability of exceeding $\chi^{2}$ for ${\nu}$ degrees of freedom
if the model is correct. \\

\clearpage
\scriptsize
\begin{center}
\begin{tabular}{lcrcccc}
\multicolumn{7}{c}{TABLE 4}\\
\multicolumn{7}{c}{Spectral Analysis of Image D of \rxj}\\
  & & & & &&\\ \hline\hline
 Observation                        &      $\Gamma$    &  $N_{\rm H}$($z_{\rm s}$=0.658)$^{a}$  & $\chi$$^{2}$/$\nu$ (cstat/$\nu$) &{\it P}($\chi$$^{2}$/$\nu$)$^{b}$&Soft Flux  & Hard Flux  \\
 Date   &    & (10$^{22}$~cm$^{-2}$)  &    & & \multicolumn{2}{c}{ (10$^{-13}$ ergs cm$^{-2}$ s$^{-1}$)} \\
\hline

2004 April 12 	&	1.95$_{-0.19}^{+0.25}$ &     0.0$_{-0.0}^{+0.10}$&    (368/789)      &   ...   &   0.67   & 0.73 \\
2006 March 10&	2.08$_{-0.47}^{+0.52}$&      0.36$_{-0.33}^{+0.40}$&    (297/837)   &   ...   &   0.81   &1.13 \\
2006 March 15	&	1.83$_{-0.44}^{+0.49}$&      0.21$_{-0.21}^{+0.32}$&    (290/837)   &  ...    &    0.78 &1.17 \\
2006 April 12	&	1.65$_{-0.25}^{+0.39}$&      0.0$_{-0.0}^{+0.26}$&    (350/837)     &      ...&    1.02&1.46 \\
2006 November 10	&	1.52$_{-0.19}^{+0.32}$&      0.0$_{-0.0}^{+0.19}$&    (403/837)       &     ...  &    1.11&1.86 \\
2006 November 13	&	2.39$_{-0.37}^{+0.53}$&       0.0$_{-0.0}^{+0.32}$&    (276/837)       &    ...   &    1.42&0.59 \\
2006 December 17       &     1.93$_{-0.36}^{+0.41}$      &    0.0$_{-0.0}^{+0.26}$  &   (347.7/2008) &    ...        &    0.83        &   1.22           \\
2007 January 01       &       1.95$_{-0.25}^{+0.27}$   &   0.0$_{-0.0}^{+0.06}$    &   (324.2/2003) &    ...    &        0.99     &     0.92        \\
2007 February 13        &       2.10$_{-0.22}^{+0.37}$   &   0.0$_{-0.0}^{+0.18}$   &       (391.9/2009)&      ...         &   1.45        &   1.23         \\
2007 February 18         &       2.01$_{-0.30}^{+0.44}$  &   0.0$_{-0.0}^{+.02}$    &      (344.2/2009) &    ...       &     1.05    &       1.14           \\
2007 April 16          &      2.35$_{-0.42}^{+0.49}$    &  0.30$_{-0.25}^{+0.30}$    &    (296.0/2009)   &    ...         &  0.79         &  0.75        \\
2007 April 25       &      1.93$_{-0.29}^{+0.37}$    &    0.0$_{-0.0}^{+0.218}$  &     (329.3/2009)   &    ...      &    1.02       &    1.27         \\
2007 June 04         &        1.79$_{-0.34}^{+0.40}$  &    0.10$_{-0.10}^{+0.27}$   &     (370.7/2003)  &    ...       &    0.80        &   1.36             \\
2007 June 11       &       2.11$_{-0.37}^{+0.41}$   &    0.243$_{-0.23}^{+0.27}$  &   (354.5/2009)  &     ...      &     0.96      &     1.21        \\
2007 July 24       &       1.77$_{-0.27}^{+0.40}$   &     0.0$_{-0.0}^{+0.22}$     &    (313.4/2009)   &     ...     &     0.71       &    1.05          \\
2007 July 30         &        2.18$_{-0.26}^{+0.47}$  &     0.0$_{-0.0}^{+0.20}$  &    (341.0/2009) &     ...      &      1.00    &      0.74          \\
2008 March 16        &         1.56$_{-0.16}^{+0.16}$    &    0.0$_{-0.0}^{+0.05}$  &    33.5/24  &   .10    &   0.83        &   1.74             \\
2008 April 13         &      1.90$_{-0.16}^{+0.21}$   &    0.043$_{-0.043}^{+0.13}$  &    20.0/26   &  .79   &    1.04       &    1.35        \\
2008 April 23        &       1.53$_{-0.21}^{+0.22}$    &   0.0$_{-0.0}^{+0.09}$   &    17.0/15   &  .32    &   0.61      &     1.23           \\
2008 June 01        &      1.77$_{-0.18}^{+0.22}$   &   0.05$_{-0.05}^{+0.14}$   &    20.6/30  &   .90   &    1.18      &     1.91              \\
2008 July 05      &        1.89$_{-0.12}^{+0.13}$ &    0.0$_{-0.0}^{+0.13}$  &      32.1/40  &   .81   &    1.76    &      2.07           \\
2008 November 11        &      1.78$_{-0.14}^{+0.16}$    &    0.0$_{-0.0}^{+0.08}$  &     25.9/29   &  .63   &    1.17   &        1.63       \\
2009 November 28       &      1.87$_{-0.05}^{+0.05}$  &   0.0$_{-0.0}^{+0.02}$  &   219.1/192  &  .09     &  5.10      &     6.25             \\
2010 February 09      &       1.94$_{-0.06}^{+0.06}$  &  0.0$_{-0.0}^{+0.03}$   &     102.4/130  &  .96   &    3.69    &       4.04       \\
2010 April 17       &     1.86$_{-0.05}^{+0.05}$    &   0.0$_{-0.0}^{+0.02}$  &      165.1/153 &  .24    &   4.14       &    5.16            \\
2010 June 25        &      1.75$_{-0.07}^{+0.10}$    &   0.0$_{-0.0}^{+0.07}$  &     76.6/91   &  .86   &    2.16    &       3.34       \\
2010 November 11       &      1.79$_{-0.06}^{+0.09}$     &    0.0$_{-0.0}^{+0.05}$ & 97.0/103  &  .65    &   2.56      &     3.60         \\
2011 January 21       &        1.68$_{-0.09}^{+0.10}$    &    0.0$_{-0.0}^{+0.10}$  &    44.3/59  &   .92    &   1.35       &    2.26           \\
2011 February 2       &      1.78$_{-0.10}^{+0.11}$      &   0.0$_{-0.0}^{+0.06}$   &        49.3/51  &   .54   &    2.17    &       3.09      \\
\hline \hline

\end{tabular}
\end{center}
${}^{a}$All model fits include fixed, Galactic absorption of $N_{\rm Gal,H}$  = 0.036 $\times$ 10$^{22}$~cm$^{-2}$ (Dickey \& Lockman 1990)
and intrinsic ($z$ = 0.658) absorption $N_{\rm H}$ set as a free parameter.
All errors are at 90\% confidence on one parameter. \\
${}^{b}$$P(\chi^2/{\nu})$ is the probability of exceeding $\chi^{2}$ for ${\nu}$ degrees of freedom
if the model is correct. The C statistic does not provide a goodness-of-fit measure. \\

\clearpage

  \begin{figure}
   \includegraphics[width=15cm]{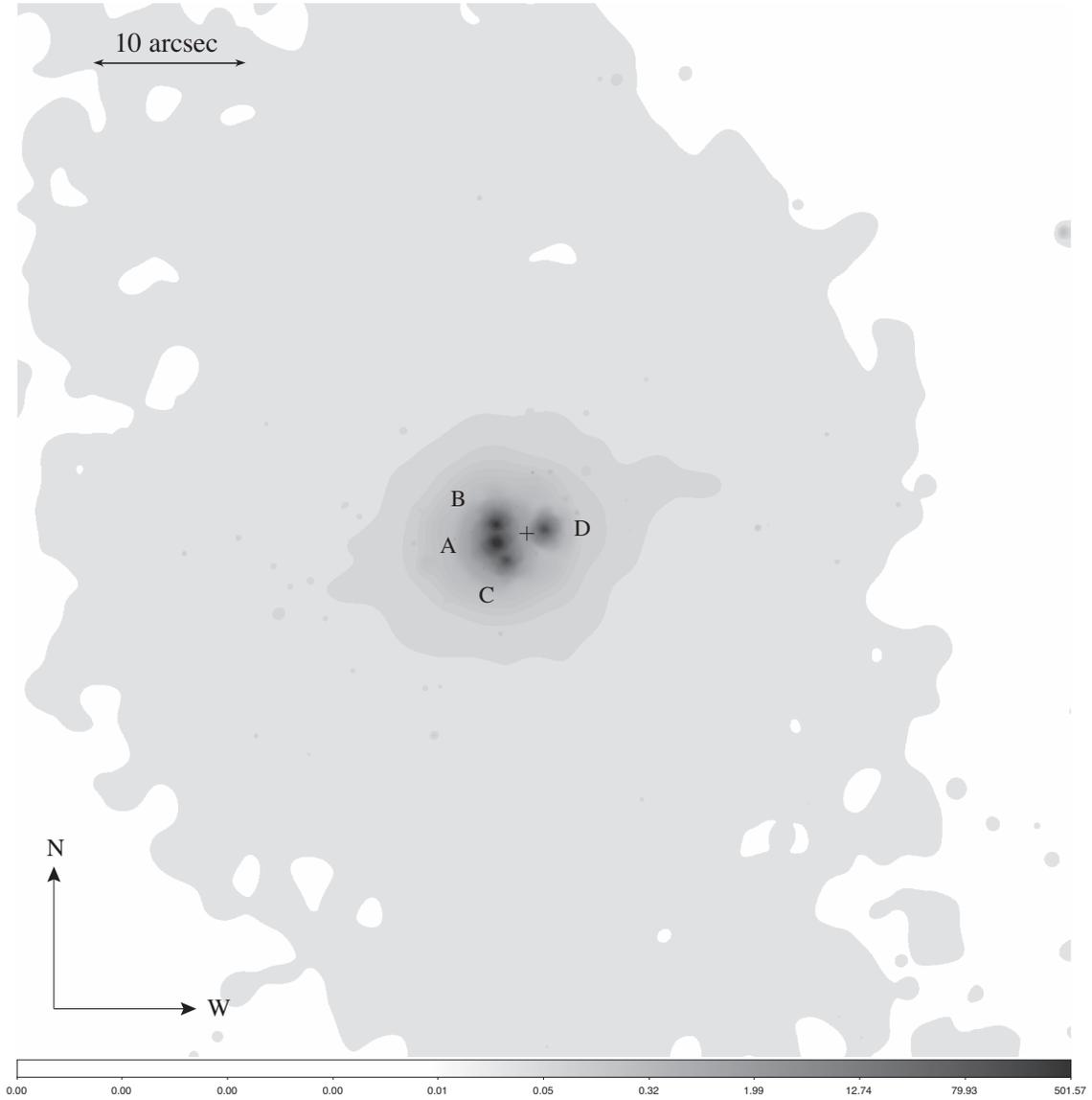}
        \centering
\caption[]{
Stacked image of all observations of \rxj\ listed in Table 1. 
The stacked image was filtered to include only photons with energies ranging between 
0.4 and 3.0 keV and adaptively smoothed.  There is no evidence for other sources in the field of 
view. The cross marks the center of the lens galaxy where we might see a fifth image.
}
\label{fig:stacked}
\end{figure}

 \begin{figure}
   \includegraphics[width=15cm]{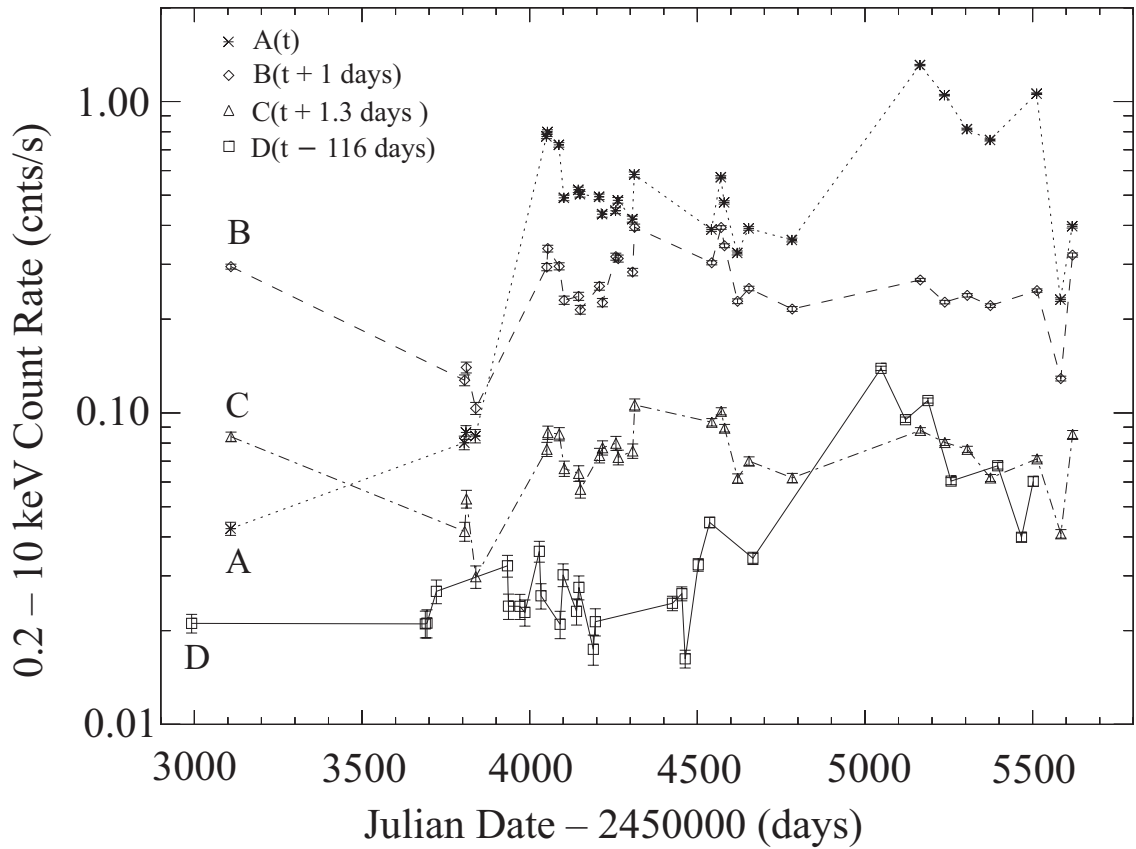}
        \centering
\caption[]{
The total (0.2$-$10~keV) light-curves of images A, B, C and D of \rxj\ shifted by the time delays estimated by 
Chantry et al. (2010).  The total counts for images A and B have been corrected for pile-up effects and pile-up 
is unimportant for images C and D. The new X-ray data begin after epoch 6 (December 2006, $\hbox{JD}-2450000=4087$). Data from epochs 1 -- 6 were presented
in Chartas et al. (2009) and Dai et al. (2010). 
}
\label{fig:lc}
\end{figure}

\begin{figure}
   \includegraphics[width=15cm]{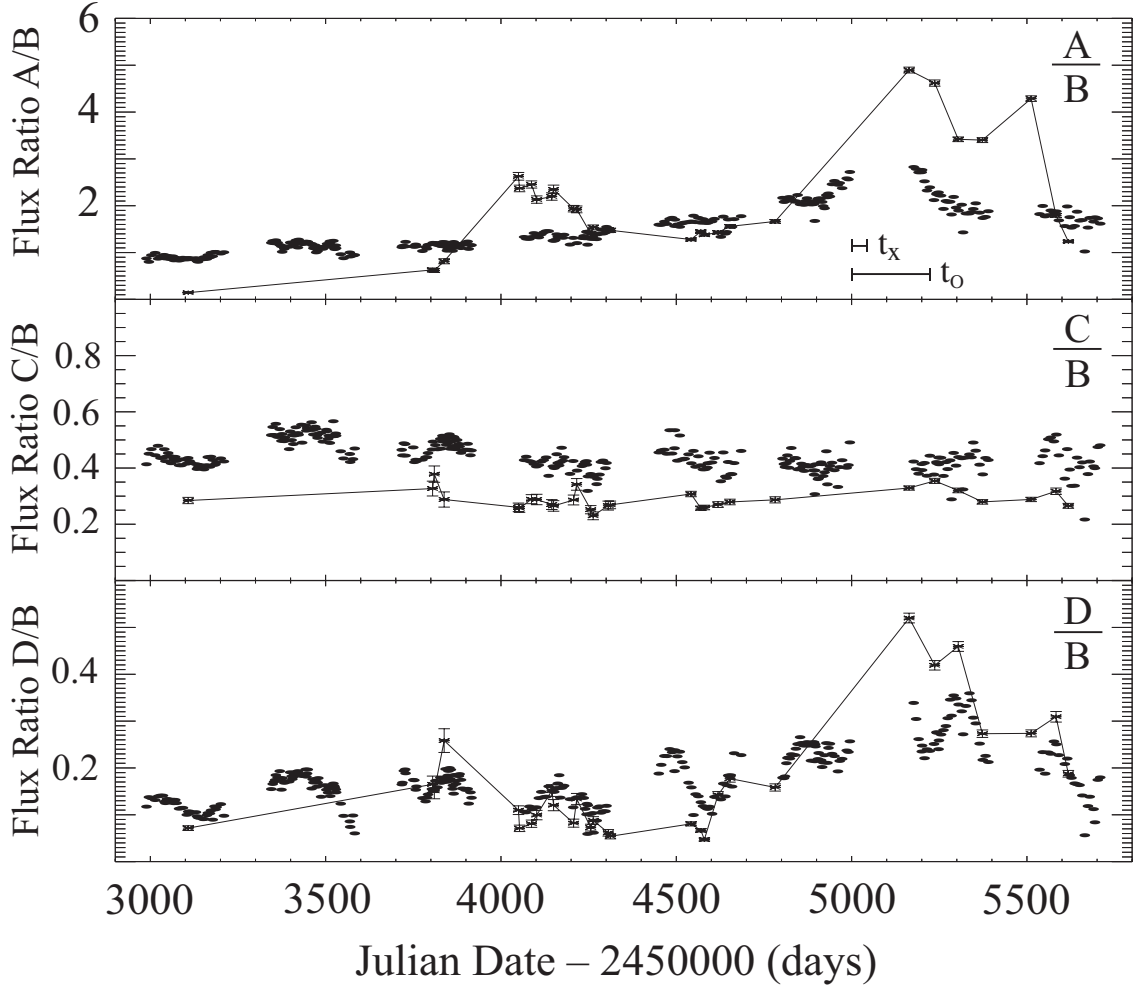}
        \centering
\caption[]{
The total (0.2$-$10~keV, crosses with error bars) and SMARTS optical ($R$-band, solid points) flux ratios A/B, C/B and D/B.  The total
band flux ratio C/B is relatively constant at 0.29 $\pm$ 0.03 supporting our hypothesis that the microlensing magnifications of 
images B and C during our observations are roughly constant.
Bars in the top panel show the estimated optical ($t_{\rm O}$) and X-ray  ($t_{\rm X}$) source crossing times for the size estimates from Dai et al. (2010)
and the effective velocity estimate from Mosquera \& Kochanek (2011a).
The new optical and X-ray data begin after day 4053. The optical and X-ray data from days 2986 -- 4053  were presented
in Chartas et al. (2009) and Dai et al. (2010). 
}
\label{fig:flux_ratios}
\end{figure}

\begin{figure}
   \includegraphics[width=15cm]{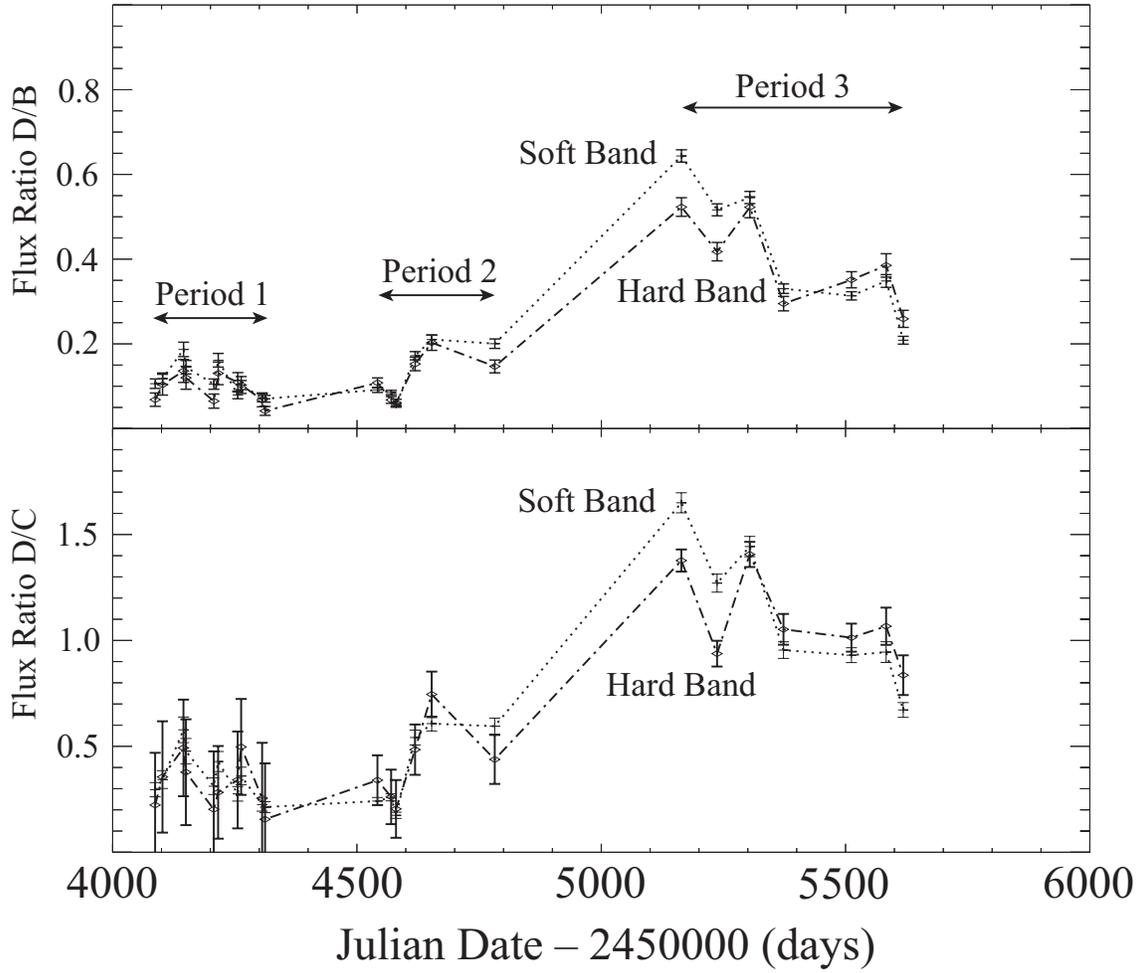}
        \centering
\caption[]{
Soft (0.2$-$2~keV) and hard (2$-$10~keV)  flux ratios D/B and D/C of \rxj.
Significant energy dependent microlensing is detected in image D.
}
\label{fig:energy_dependent_ml}
\end{figure}

\begin{figure}
   \includegraphics[width=15cm]{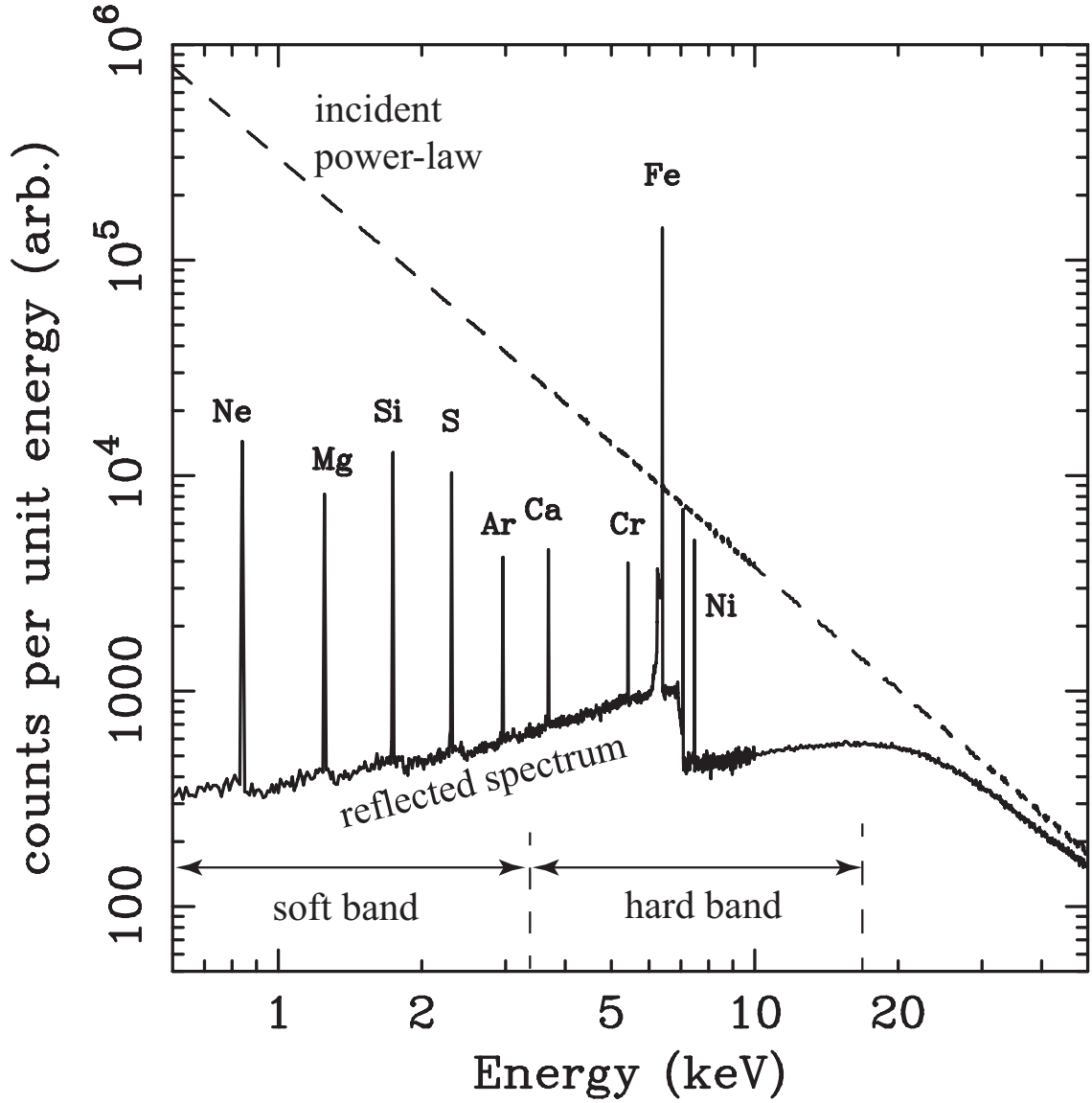}
        \centering
\caption[]{
Simulated reflection spectrum from a cold accretion disk following Reynolds et al. (1998). The dashed line indicates the 
input incident power-law spectrum of the corona. We have over-plotted the rest-frame soft and hard bands
used in our analysis. The soft band contains a larger fraction of direct emission than the hard band. 
The dominant fluorescence line is Fe~K$\alpha$ at 6.4~keV.
}
\label{fig:reflection}
\end{figure}

\begin{figure}
      \includegraphics[width=15cm]{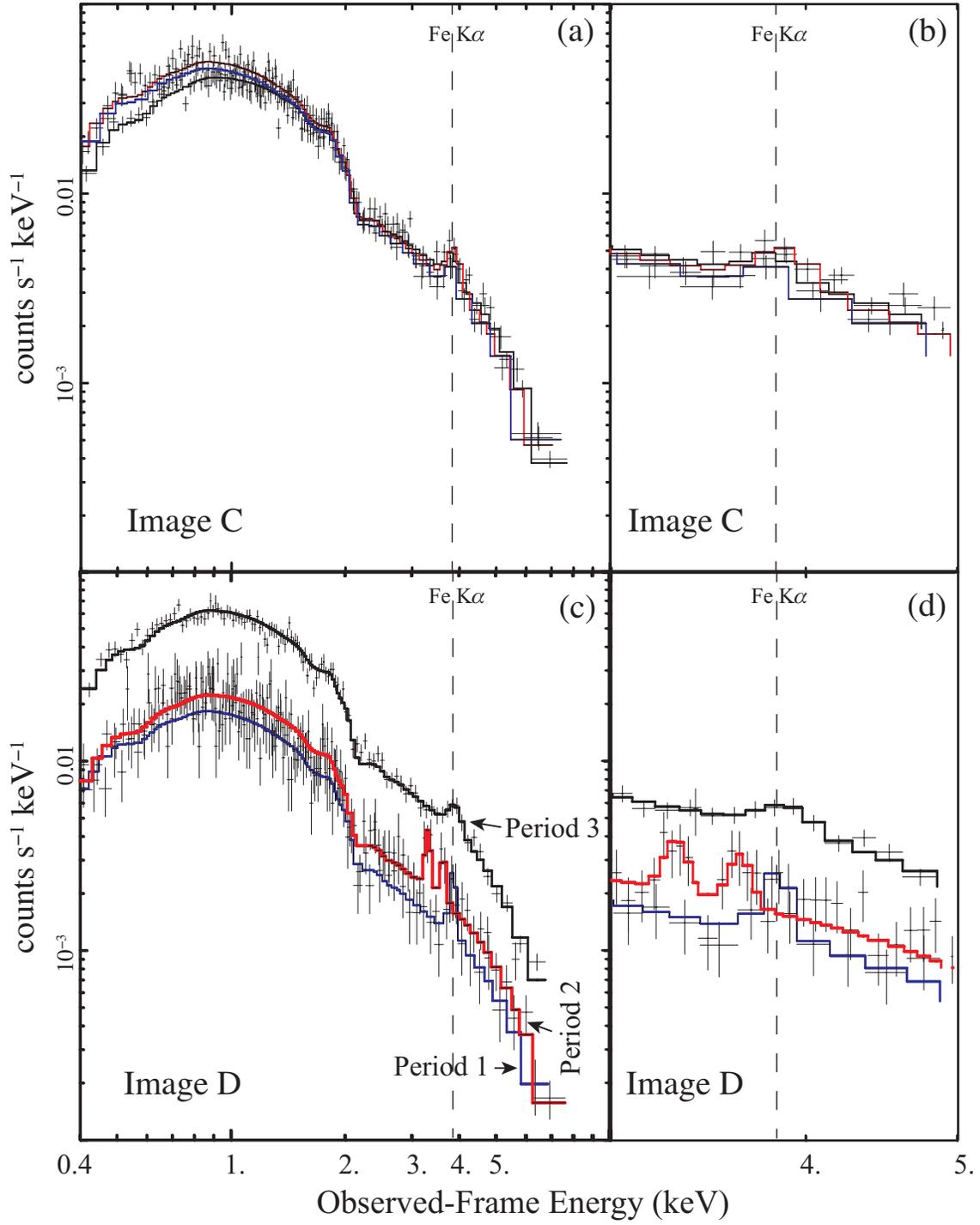}
        \centering
\caption[]{
Stacked spectra of images C (panel a) and D (panels c) for the three periods shown in Figure 4.
A close-up of the stacked spectra around the Fe~K$\alpha$ region for images C (panel b) and D (panel d).
The curves are the best-fit absorbed power-law plus Gaussian line models.
}
\label{fig:double_peaked_line}
\end{figure}

\begin{figure}
      \includegraphics[width=15cm]{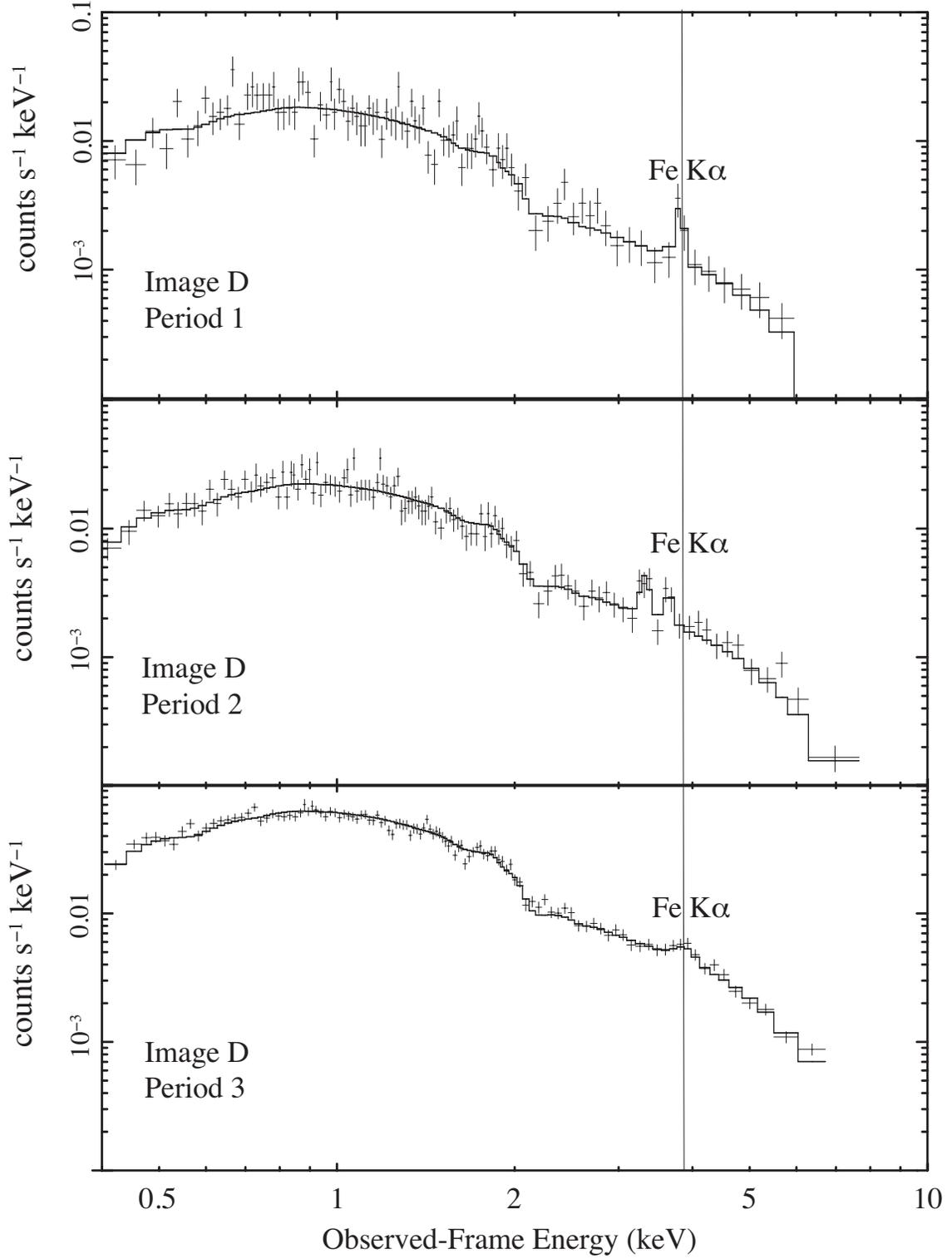}
        \centering
\caption[]{
Stacked spectra of image D shown in separate panels for the three periods shown in Figure 4.
The curves are the best-fit absorbed power-law plus Gaussian line models.
}
\label{fig:fe_line_evolution}
\end{figure}

\begin{figure}
   \includegraphics[width=15cm]{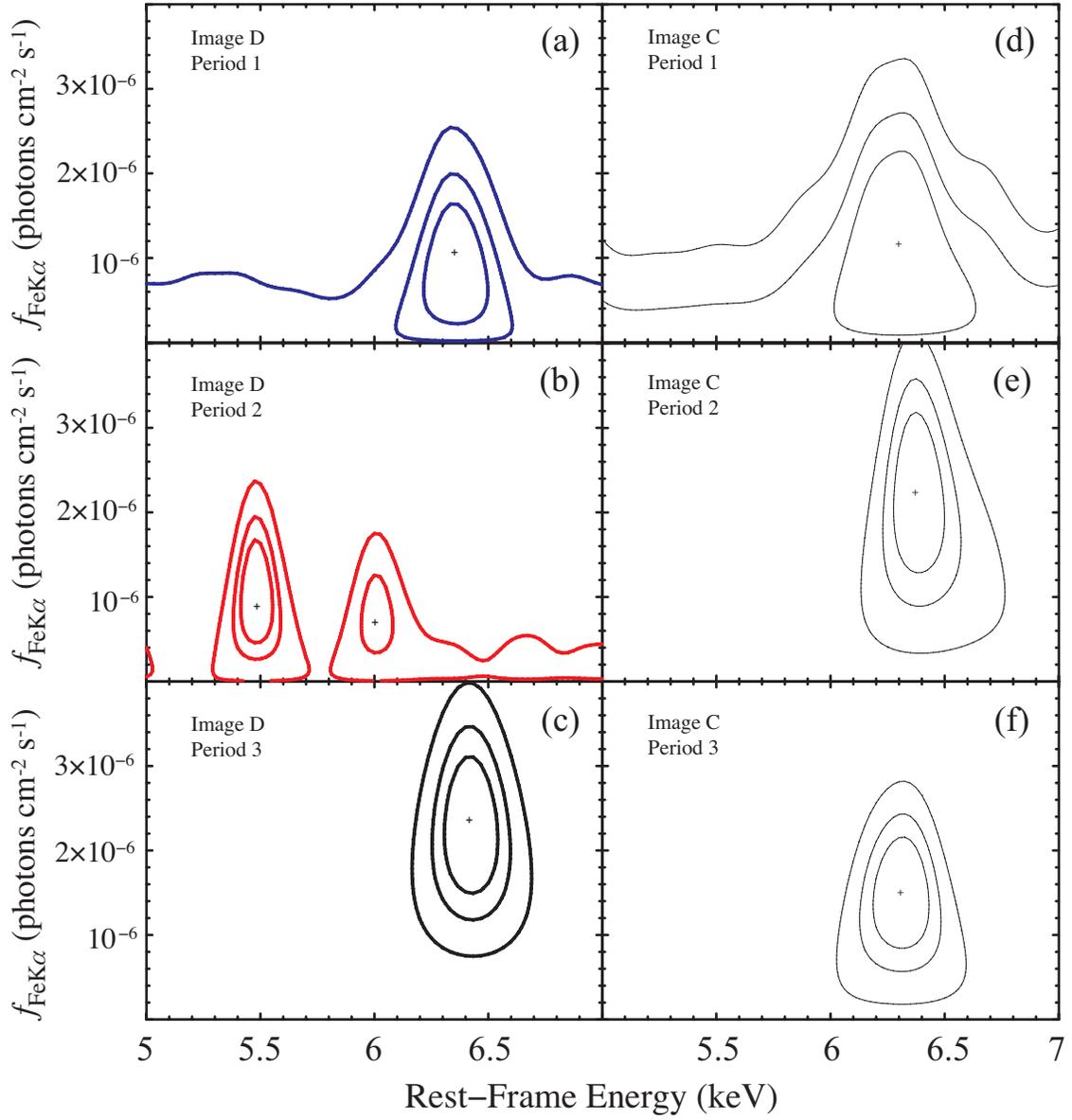}
        \centering
\caption[]{
The contours represent 68\%, 90\% and 99\% $\chi^{2}$ confidence intervals of the flux normalizations, $f_{\rm Fe~K\alpha}$, of the detected 
Fe K$\alpha$ line in image D (left panels) and in image C (right panels) in periods 1 (panels a and d), 2 (panels b and e) and 3 (panels c and f) from Figure 4.  
}
\label{fig:confidence_contours}
\end{figure}

\begin{figure}
   \includegraphics[width=15cm]{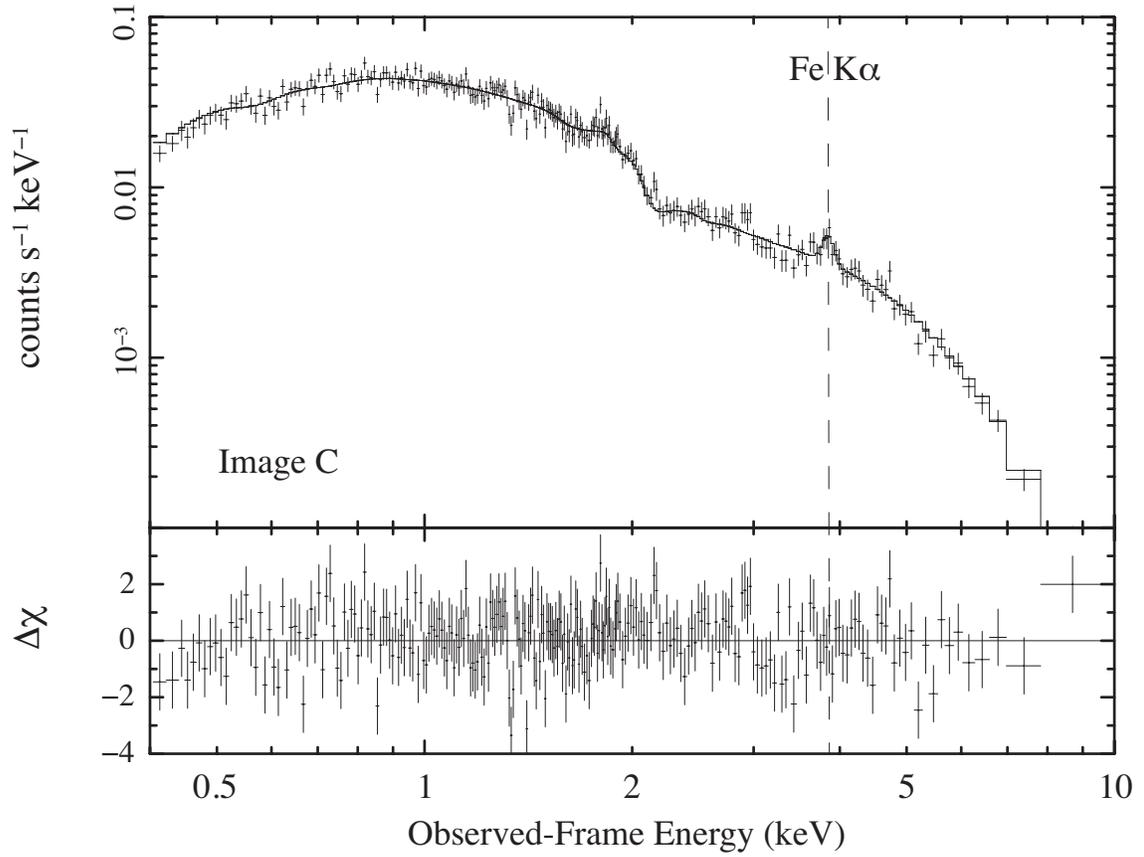}
       \centering
\caption[]{
The top panel shows the stacked spectrum  of image C for all epochs fit by a power-law plus a Gaussian line model with Galactic absorption.
The vertical dashed line indicates the energy of an Fe-K$\alpha$ line with a rest-frame energy of 6.4~keV.
The lower panel shows the residuals of the fit in units of 1$\sigma$ deviations.
}
\label{fig:image_C_stacked}
\end{figure}

\begin{figure}
   \includegraphics[width=13cm]{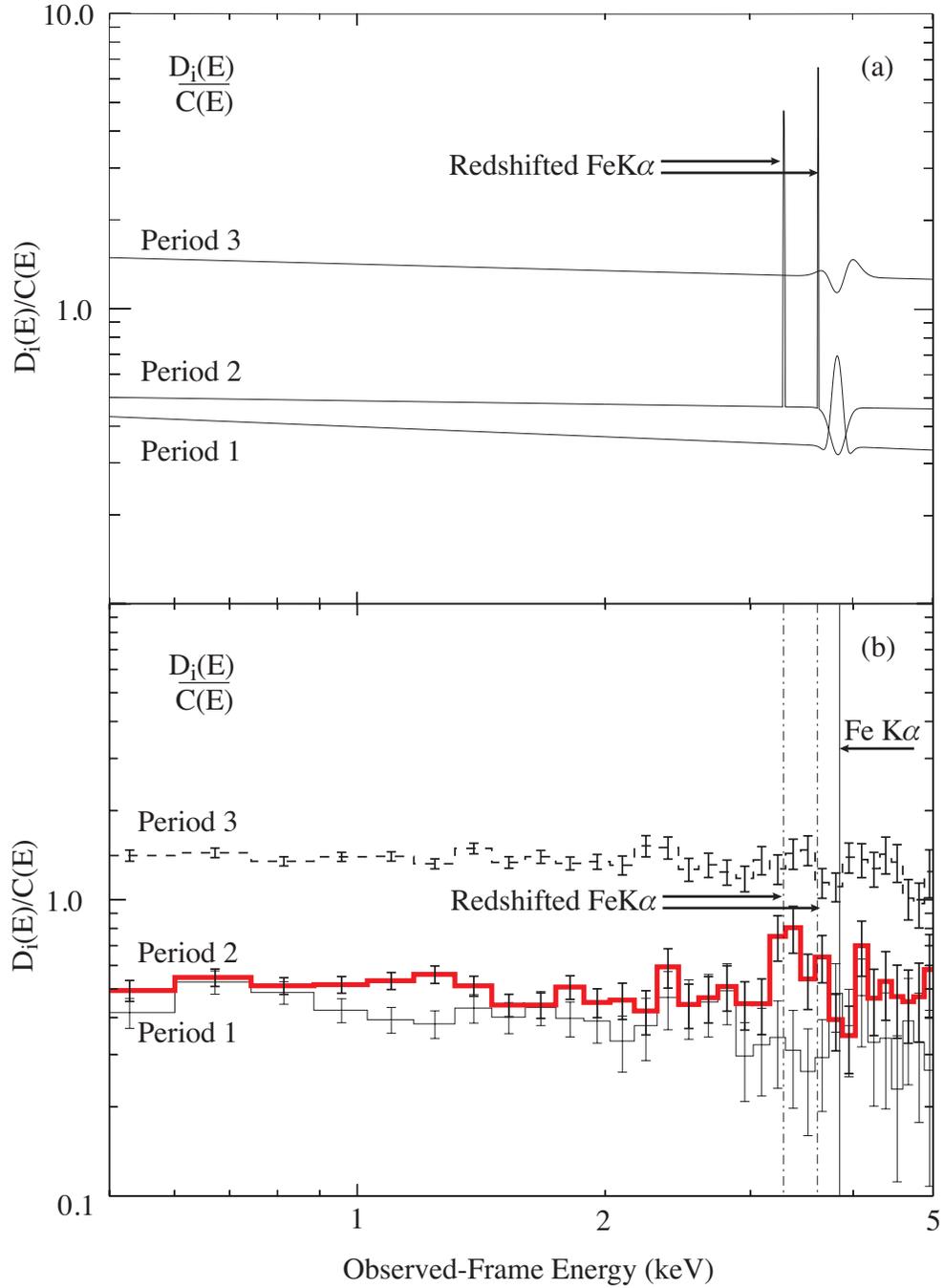}
       \centering
\caption[]{
Ratios of the (a) best-fit models (shown in Figure 6) and (b) observed flux densities 
$D_{\rm i}(E)/C(E)$, where, 
$D_{\rm i}(E)$ is the flux density for image D in period $i$ and $C(E)$ is the flux density  for
image C in all three periods. The three periods are shown in Figure 4.
The model spectra are much sharper because they are not convolved to the instrumental resolution.
}
\label{fig:ratiosD_C}
\end{figure}

\end{document}